\documentclass[
 aip,
 amsmath,amssymb,
 reprint
]{revtex4-2}

\usepackage[utf8]{inputenc}
\usepackage{graphicx}
\usepackage{verbatim}
\usepackage{amsmath}
\usepackage{amssymb}
\usepackage{dsfont}
\usepackage{float}
\usepackage{cancel}
\usepackage{color}
\usepackage[usenames,dvipsnames,svgnames,table]{xcolor}
\usepackage{comment}
\usepackage{ulem}
\usepackage{stackengine}
\usepackage{hyperref}
\usepackage{csquotes}
\usepackage{booktabs}
\usepackage{bm}
\usepackage[T1]{fontenc}
\usepackage{mathptmx}
\usepackage{etoolbox}
\usepackage{nicefrac}
\usepackage{empheq}
\newcommand*\widefbox[1]{\fbox{\hspace{1em}#1\hspace{1em}}}

\newcommand{\vect}[1]{\overrightarrow{#1}}
\newcommand{\mat}[1]{\pmb{#1}}

\newcommand{\Tr}{\operatorname{Tr}}

\newcommand{\comm}[2]{\left[#1, #2\right]}

\newcommand{\lrangle}[1]{\big\langle #1 \big\rangle}
\newcommand{\op}[1]{\hat{#1}}
\newcommand{\supop}[1]{\mathbb{#1}}
\newcommand{\lvl}{\supop{L}}

\newcommand{\supopI}{\mathds{1}}
\newcommand{\nnref}[3]{\hyperref[#1]{#2\ref*{#1}#3}}

\makeatletter
\def\@email#1#2{%
 \endgroup
 \patchcmd{\titleblock@produce}
  {\frontmatter@RRAPformat}
  {\frontmatter@RRAPformat{\produce@RRAP{*#1\href{mailto:#2}{#2}}}\frontmatter@RRAPformat}
  {}{}
}%
\makeatother

\renewcommand{\Dot}[1]{\overset{\footnotesize \bullet}{#1}}

\begin{document}

\preprint{AIP/123-QED}

\title{Note: Nonuniqueness of generalized quantum master equations for a single observable}
\author{Nathan Ng}
 \email{nang@berkeley.edu}
 \affiliation{Department of Physics, University of California, Berkeley, CA 94720, USA}
\author{David T. Limmer}%
\affiliation{ 
Department of Chemistry, University of California, Berkeley, CA 94720, USA
}
 \affiliation{Kavli Energy NanoScience Insitute, University of California, Berkeley, CA 94720, USA}
\affiliation{Materials Sciences Division, Lawrence Berkeley National Laboratory, Berkeley, CA 94720,~USA \looseness=-1}
\affiliation{Chemical Sciences Division, Lawrence Berkeley National Laboratory, Berkeley, CA 94720,~USA \looseness=-1}

\author{Eran Rabani}
\affiliation{ 
Department of Chemistry, University of California, Berkeley, CA 94720, USA 
}
\affiliation{Materials Sciences Division, Lawrence Berkeley National Laboratory, Berkeley, CA 94720,~USA \looseness=-1}
\affiliation{The Sackler Center for Computational Molecular and Materials Science, Tel Aviv University, Tel Aviv 69978,~Israel \looseness=-1}

\date{\today}

\begin{abstract}
When deriving exact generalized master equations for the evolution of a reduced set of degrees of freedom, one is free to choose what quantities are relevant by specifying projection operators. 
However, obtaining a reduced description does not always need to be achieved through projections--one can also use conservation laws for this purpose.
Such an operation should be considered as distinct from any kind of projection; that is, projection onto a single observable yields a different form of master equation compared to that resulting from a projection followed by the application of a constraint.
We give a simple example to show this point and give relationships that the different memory kernels must satisfy to yield the same dynamics.
\end{abstract}

\maketitle

In the study of the dynamics of large, closed quantum systems, an approach often taken is to model the dynamics of a reduced set of degrees of freedom (dofs) using the exact Nakajima-Zwanzig (NZ) equation (in units where $\hbar = 1$),\cite{Nakajima1958,Zwanzig1960}
\begin{align}
    \label{eq:nakajima-zwanzig}
    \begin{split}
    \frac{d}{dt} \supop{P} \op{\rho}(t) &= -i \left( \supop{P} \lvl \supop{P} \right) \supop{P} \op{\rho}(t) + \op{\theta}(t)  \\ &\quad- \int_0^t d\tau \, \supop{K}(\tau; \supop{P}) \,\, \supop{P} \op{\rho}(t - \tau).
\end{split}
\end{align}
In the above equation, $\op{\theta}(t) = - i \supop{P} \lvl e^{-i t \supop{Q} \lvl} \supop{Q} \op{\rho}(0)$ is the inhomogeneous term and the memory kernel superoperator for an arbitrary projection $\supop{P}$ is given by,
\begin{equation}
\supop{K}(\tau; \supop{P}) = \supop{P} \lvl e^{-i \tau \supop{Q} \lvl} \supop{Q} \lvl \supop{P},
\label{eq:memory}
\end{equation}
where $\lvl \, \cdots = \comm{\op{H}}{\,\cdots \,}$ is the Liouvillian superoperator, with $\op{H}$ being the Hamiltonian of the entire closed quantum system.
The NZ equation is derived from the quantum Liouville equation by making use of projection superoperators $\supop{P}$ and its complement $\supop{Q} \equiv \supopI - \supop{P}$ acting on the full density matrix $\op{\rho}(t)$.
There are no restrictions on the choice of $\supop{P}$, so different choices will alter \nnref{eq:nakajima-zwanzig}{Eqs.~(}{)} and \nnref{eq:memory}{(}{)} and give rise to different quantum master equations for the same observable which, in principle, should yield the same dynamics for this quantity.

In this note we will consider a system coupled to a ``bath'' and derive two apparently different quantum master equations for one of the system's populations.\footnote{We put ``bath'' in quotations since we are using it as a shorthand for degrees of freedom we do not care about, rather than it being an infinitely large reservoir that is unchanged by coupling to a small system.}
This example illustrates how two structurally distinct master equations, one that contains a term acting like an external drive and one that does not, can result from our choice of $\supop{P}$ and whether we impose conservation laws on the set of reduced variables.
For simplicity we will restrict our discussion to a two-level system (TLS) coupled to some number of other dofs which we call the bath $B$. 
The general case of an $D$-level system coupled to other dofs is described in section III of the Supplementary Material.~\footnote{Details can be found in the Supplementary Material at URL.}
%Operators acting on the full Hilbert space are generally formed from sums of factorized terms, $\op{S}\otimes\op{B}$, where $\op{S}$ and $\op{B}$  act only on the TLS and the bath, respectively.
Throughout this note, we take a factorized initial condition, $\op{\rho}(0) = \supop{P}\op{\rho}(0)$, so that $\supop{Q}\op{\rho}(0) = 0$, and $\op{\theta}(t) = 0$.

We will now specify the two projectors used to derive two quantum master equations.
For notational convenience, we define the projectors on to the $n$th population of the system for some arbitrary reference bath state $\op{\rho}_B$,
\begin{align}
    \label{eq:pn-projectors}
    \supop{P}^n \op{\rho} &= \bigg( |n\rangle\langle n| \otimes \op{\rho}_B \bigg) \Tr \bigg\{ \left( |n\rangle\langle n| \otimes \op{I}_B \right) \op{\rho} \bigg\}.
\end{align}
In the first case, we take $\supop{P} = \supop{P}^0 + \supop{P}^1$, which results in a set of coupled master equations for the populations $|0\rangle\langle 0|$ and $|1\rangle\langle 1|$.
From here, we can reduce the description even further since the system obeys $\sigma_0(t) + \sigma_1(t) = 1$, and focus solely on the $|0\rangle\langle 0|$ component.
In the second case, we consider $\supop{P} = \supop{P}^0$, and therefore deem the $|0\rangle\langle 0|$ component of the reduced density matrix, $\sigma_{0}(t) \equiv \rho_{00}(t)$, to be the \textit{only} component of interest.
We shall now show how these two procedures lead to structurally distinct generalized quantum master equations, and how the associated memory kernels must relate to each other so that they produce the same dynamics for $\sigma_{0}(t)$.

For the first case, we use $\supop{P} = \supop{P}^0 + \supop{P}^1$ and $\supop{Q} = \supopI - \supop{P} = \supop{Q}^0 - \supop{P}^1$ in \nnref{eq:nakajima-zwanzig}{Eqs.~(}{)} and \nnref{eq:memory}{(}{)} to obtain,
\begin{align}
    \label{eq:p0-p1-nzme}
    \negthickspace\frac{d\sigma_{0}(t)}{dt} &= \int_0^t \!\!d\tau \, K_{1,1}(\tau; \supop{P}^0 + \supop{P}^1) \\ &\quad - \bigg[ K_{0,0}(\tau; \supop{P}^0 + \supop{P}^1) + K_{1,1}(\tau; \supop{P}^0 + \supop{P}^1)\bigg] \, \sigma_{0}(t-\tau), \nonumber
\end{align}
% with the memory kernels given by:
% \begin{equation}
%     K_{m,n}(\tau; \supop{P}) = \Tr\bigg[ (|m\rangle\langle m| \otimes \op{I}_B) \lvl e^{-i \tau \supop{Q} \lvl} \supop{Q} \lvl (|n\rangle\langle n| \otimes \op{\rho}_B) \bigg].\!\!\label{eq:p0-p1-memory}
% \end{equation}
We use the notation $K_{m,n}(\tau; \supop{P})$ to denote the $(m,n)$ matrix element of the memory kernel superoperator $\supop{K}$ in \nnref{eq:memory}{Eq.~(}{)} (see Ref. [\onlinecite{Note2}]).
In obtaining \nnref{eq:p0-p1-nzme}{Eq.~(}{)} we have also used the relations $\sigma_0(t) + \sigma_1(t) = 1$ and $K_{0,n}(t) + K_{1,n}(t) = 0$.
%Note that $K^{(0)}(\tau) \ne K_{0,0}(\tau)$ since $\supop{Q}^0 \ne \supop{Q}$ in \nnref{eq:k0-memory}{Eqs.~(}{)} and \nnref{eq:p0-p1-memory}{(}{)}.

For the second case, using $\supop{P} = \supop{P}^0$ and $\supop{Q} = \supopI - \supop{P}^0$ in \nnref{eq:nakajima-zwanzig}{Eqs.~(}{)} and \nnref{eq:memory}{(}{)}, we obtain,
\begin{equation}
    \label{eq:p0-nzme}
    \frac{d\sigma_{0}(t)}{dt} = - \int_0^t d\tau \, K_{0,0}(\tau; \supop{P}^0) \, \sigma_{0}(t-\tau).
\end{equation}
%with the (scalar) memory kernel given by:
% \begin{equation}    
% K^{(0)}(\tau) = \Tr\bigg[ (|0\rangle\langle 0| \otimes \op{I}_B) \lvl e^{-i \tau \supop{Q}^0 \lvl} \supop{Q}^0 \lvl (|0\rangle\langle 0| \otimes \op{\rho}_B) \bigg].\!\!\! \label{eq:k0-memory}
% \end{equation}

\nnref{eq:p0-p1-nzme}{Eq.~(}{)} superficially differs from \nnref{eq:p0-nzme}{Eq.~(}{)} in that former contains a term independent of the population at time $t$ and thus plays a role like an external force in a Langevin equation.
However, this force arises due to a need to satisfy a constraint, $\sigma_0(t) + \sigma_1(t) = 1$, which used to be enforced by properties of the memory kernel, $K_{0,n}(t) + K_{1,n}(t) = 0$.
%By projecting only on to the relevant dof at the outset to obtain \nnref{eq:p0-nzme}{Eq.~(}{)}, we will therefore not incur an inhomogeneous term; rather, 
%We will formalize this intuition by relating this term to the known memory kernels $K^0(\tau)$ to prove the equivalence of the dynamics generated by \nnref{eq:p0-p1-nzme}{Eqs.~(}{)} and \nnref{eq:p0-nzme}{(}{)}.

\begin{figure}
    \centering
    \includegraphics[width=\columnwidth]{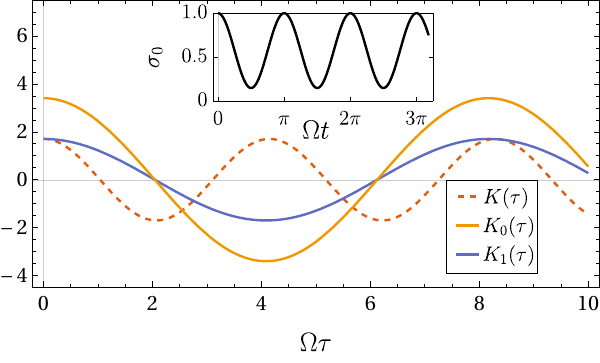}
    \caption{(Color online) Memory kernels for the two-level system using the two projection schemes resulting in master equations \nnref{eq:p0-p1-nzme}{Eqs.~(}{)}-\nnref{eq:p0-nzme}{(}{)}. The initial condition is chosen as $\op{\rho}(0) = |0\rangle\langle 0|$, and the parameters are $\Delta = 12/13$ and $\varepsilon = 5/13$. Both master equations generate the same dynamics for the population $\sigma_0(t)$ (inset). For brevity, we use $K(\tau) \equiv K_{0,0}(\tau; \supop{P}^0)$, $K_1(\tau) \equiv K_{1,1}(\tau; \supop{P}^0 + \supop{P}^1)$, and $K_0(\tau) \equiv K_{0,0}(\tau; \supop{P}^0 + \supop{P}^1) + K_{1,1}(\tau; \supop{P}^0 + \supop{P}^1)$.
    }
    \label{fig:qubit-memory-comparison}
\end{figure}

While our preceding discussion has been fully general albeit abstract, let us be more concrete by working with the simple example of a TLS with states $|0\rangle$ and $|1\rangle$ not coupled to any bath.
The system is described by the Hamiltonian $\op{H} = \varepsilon \op{\tau}^z + \Delta \op{\tau}^x$, where $\op{\tau}^\alpha$ ($\alpha=x,z$) are Pauli matrices. 
For the first case as outlined above, we use $\supop{P} \cdots = \supop{P}^n\cdots =( |n\rangle\langle n|) \Tr \{ \left( |n\rangle\langle n| \right) \cdots \}$ for $n=0,1$ and the initial condition $\op{\rho}(0) = |0\rangle\langle 0|$ to get explicit memory kernels (see Ref. [\onlinecite{Note2}]),
\begin{align}
    \label{eq:tls-p0-p1-memory}
    K_{0,0}(\tau; \supop{P}^0 + \supop{P}^1) = K_{1,1}(\tau; \supop{P}^0 + \supop{P}^1) = 2 \Delta^2 \cos(2 \varepsilon \tau).
\end{align}
For the second case, we obtain instead,
\begin{align}
    \label{eq:tls-p0-memory}
    K_{0,0}(\tau; \supop{P}^0) = 2 \Delta^2 \cos(2 \omega \tau), \qquad \omega = \sqrt{\varepsilon^2 + (1/2)\Delta^2}.
\end{align}
These memory kernels are quite distinct, as shown in \nnref{fig:qubit-memory-comparison}{Fig.~}{}, but the population dynamics they generate are identical. 
It is straightforward to compute $\op{\rho}(t) = e^{-i \op{H} t} \op{\rho}(0) e^{i \op{H} t}$ for the above model Hamiltonian and verify that the following function solves \nnref{eq:p0-p1-nzme}{Eqs.~(}{)} and \nnref{eq:p0-nzme}{(}{)} when $\sigma_0(0) = 1$:
\begin{align}
    \label{eq:tls-solution}
\sigma_{0}(t) = 1 - \frac{\Delta^2 \sin^2(\Omega t)}{\Omega^2}, \qquad \Omega = \sqrt{\varepsilon^2 + \Delta^2}.
\end{align}

Equivalence of the dynamics between \nnref{eq:p0-p1-nzme}{Eqs.~(}{)} and \nnref{eq:p0-nzme}{(}{)} should imply relations between the memory kernels in the two equations.
It is convenient to work with $F_{m,m}(\tau; \supop{P})$ such that its time-derivative generates the memory $K_{m,m}(\tau; \supop{P})$,
\begin{align}
    F_{m,m}(\tau; \supop{P}) = \Tr\bigg[ (|m\rangle\langle m| \otimes \op{I}_B) \lvl e^{-i \tau \supop{Q} \lvl} (|m\rangle\langle m| \otimes \op{\rho}_B) \bigg],\!\!\!\label{eq:f-def}
\end{align}
where $m = 0,1$.
The condition that the functions $F(t)$ with respect to different projectors must generate the same $p_0(t)$ is equivalent to the relation:
\begin{align}
    \label{eq:f0-f-identity}
    F_{m,m}(t; \supop{P}^0) &= F_{m,m}(t; \supop{P}^0 + \supop{P}^1) \\ 
    &\quad - \int_0^t d\tau \, F_{1,1}(\tau; \supop{P}^0) \, F_{m,m}(t - \tau; \supop{P}^0 + \supop{P}^1). \nonumber 
\end{align}
This identity is easily demonstrated by working with $\widetilde{F}$, the Laplace transform of $F$, and applying Dyson's identity, $(X - Y)^{-1} = X^{-1} + X^{-1} Y (X - Y)^{-1}$.
Using \nnref{eq:f0-f-identity}{Eq.~(}{)} in Laplace space, it is straightforward to show that the $\sigma_0(t)$ solving \nnref{eq:p0-p1-nzme}{Eq.~(}{)},
\begin{align}
    \widetilde{\sigma}_0(z) &= \frac{1}{z} \left[ \frac{1 + \widetilde{F}_{1,1}(z; \supop{P}^0 + \supop{P}^1)}{1 + \widetilde{F}_{0,0}(z; \supop{P}^0 + \supop{P}^1) + \widetilde{F}_{1,1}(z; \supop{P}^0 + \supop{P}^1)} \right],
\end{align}
is identical to the one solving \nnref{eq:p0-nzme}{Eq.~(}{)}, see Ref.~[\onlinecite{Note2}].

We can turn back to the TLS and note that the functions
\begin{align}
    \begin{split}
        F_{0,0}(\tau; \supop{P}^0 + \supop{P}^1) &= F_{1,1}(\tau; \supop{P}^0 + \supop{P}^1) = \left(\frac{\Delta^2}{\varepsilon} \right) \sin(2\varepsilon \tau) \\
        F_{0,0}(\tau; \supop{P}^0) =&\,\, F_{1,1}(\tau; \supop{P}^0) = \left(\frac{\Delta^2}{\omega} \right) \sin(2\omega \tau),
    \end{split}
\end{align}
indeed satisfy \nnref{eq:f0-f-identity}{Eq.~(}{)}.

The structural difference between the master equations--arising from projections and conservation laws--is not limited to a system of two states.
While our point is most striking for a TLS, differences can also appear for general $D$-level systems whenever the conservation of total population allows for the reduction of one additional dof.
It is important to remember that the memory kernels associated with different projectors are not independent from each other.
While time-nonlocal and complicated, relations like \nnref{eq:f0-f-identity}{Eq.~(}{)} show that the memory kernels contain information in common about the underlying unitary dynamics.
This information comes from conservation laws, the presence of which is explicitly ignored by the projection operations.
Thus, our intuition is that different evolution equations can arise depending whether one imposes constraints or projects out dofs (see Ref.~[\onlinecite{Note2}]).
Thus, there is no guarantee of a unique generalized quantum master equation to describe the reduced dynamics of an observable.
We note that a related observation have been made about Markovian quantum master equations.~\cite{Levante1995} \\

{\noindent}{\bf Supplementary Material}: 
See supplementary material for 1) a quick derivation of the Nakajima-Zwanzig equation with generic projection superoperators; 2) a general demonstration of how inhomogeneous terms in the Nakajima-Zwanzig equation appear if conservation laws are explicitly imposed; 3) a generalization of the time-nonlocal relation \nnref{eq:f0-f-identity}{Eq.~(}{)} for $D$-level systems; and 4) details on the two-level system example.
\\

{\noindent}{\bf Acknowledgements}: This work was supported by the U.S. Department of Energy, Office of Science, Office of Basic Energy Sciences, Materials Sciences and Engineering Division, under Contract No. DEAC02-05-CH11231, within the Physical Chemistry of Inorganic Nanostructures Program (KC3103). The codes that support this study are available from  upon reasonable request.  \\

{\noindent}{\bf Author Declarations}: 

{\bf Conflicts of Interest:} The authors have no conflicts to disclose.

\bibliography{references}
\clearpage
\newpage
\setcounter{page}{1}
\setcounter{equation}{0}
\onecolumngrid

\centerline{\Large Supplementary Material: Nonuniqueness of generalized master equations for a single observable}

\vspace*{2em}

This supplement is divided into four parts:
\begin{itemize}
    \item[\nnref{sec:SM-nakajima-zwanzig}{SM~}{}: ] We review a general derivation of Nakajima-Zwanzig equations for general choices of projection superoperators using only the assumption of an autonomous, first order linear homogeneous dynamical system. 
    \item[\nnref{sec:SM-constraints}{SM~}{}: ] We show generically how structurally distinct Nakajima-Zwanzig master equations can arise from imposing dynamical constraints in the form of conserved charges.
    \item[\nnref{sec:SM-dynamics-equivalence}{SM~}{}: ] We show how to prove equivalence of dynamics between the different master equations presented in the main text, and generalize the time-nonlocal relation between memory kernels (\nnref{eq:f0-f-identity}{Eq.~(}{)} of the main text) to $D$-state systems.
    \item[\nnref{sec:SM-tls}{SM~}{}: ] We give details on the two-level system example stated in the main text, deriving ``master equations'' and their associated memory kernels from four different approaches to obtaining reduced dynamics. These four approaches differ in how one imposes dynamical constraints and uses the Nakajima-Zwanzig equation for reduced dynamics.
\end{itemize}

\section{\label{sec:SM-nakajima-zwanzig}Derivation of the Nakajima-Zwanzig equation}
%In this section, we give a quick derivation of the Nakajima-Zwanzig equation, keeping the choice of projection superoperator $\supop{P}$ completely general.

In a closed system, time evolution in the Schr\"odinger picture is governed by the Liouville-von Neumann equation,
\begin{align}
\label{eq:SM-liouville-eq}
\frac{d}{dt} \op{\rho} &= -i \lvl \op{\rho} \equiv -i \comm{\op{H}}{\op{\rho}}.
\end{align}
Laplace transforming both sides yields
\begin{align}
\label{eq:SM-liouville-laplace}
  z \op{\widetilde{\rho}}(z) - \op{\rho}(0) &= -i \lvl \op{\widetilde{\rho}}(z) \nonumber \\
  \op{\widetilde{\rho}}(z) &= \frac{1}{z + i \lvl} \op{\rho}(0) \nonumber \\
  &= \frac{1}{z + i \lvl \supop{Q} + i \lvl \supop{P}} \op{\rho}(0),
\end{align}
where in the last line we used the fact that $\supopI = \supop{P} + \supop{Q}$.
Working in Laplace space is convenient, partly because it circumvents the need to deal with manipulating integrals.
For instance, we can directly interpret the product of two Laplace-transformed functions as a convolution.
We proceed via Dyson's identity in algebraic form,
\begin{align}
\label{eq:SM-dyson-identity}
(X \pm Y)^{-1} &= X^{-1} \mp X^{-1} Y (X \pm Y)^{-1},
\end{align}
where $X$ and $(X\pm Y)$ are assumed to be invertible.
Applying this twice on \nnref{eq:SM-liouville-laplace}{Eq.~(}{)} gives
\begin{align*}
  \op{\widetilde{\rho}}(z) &= \frac{1}{z + i \lvl \supop{Q}} \op{\rho}(0) - i \frac{1}{z + i \lvl \supop{Q}} \lvl \supop{P} \frac{1}{z + i \lvl \supop{Q} + i \lvl \supop{P}} \op{\rho}(0) \\
                           &= \frac{1}{z} \op{\rho}(0) - \frac{i}{z} \lvl \supop{Q} \frac{1}{z + i \lvl \supop{Q}} \op{\rho}(0) - i \frac{1}{z + i \lvl \supop{Q}} \lvl \supop{P} \op{\widetilde{\rho}}(z),
\end{align*}
where in the last line we have applied Dyson's identity to the first term on the RHS of the first line. 
To arrive back at a differential equation, we multiply both sides by $z$ and subtract $\op{\rho}(0)$:
\begin{align*}
  z \op{\widetilde{\rho}}(z) - \op{\rho}(0) &= - i \lvl \supop{Q} \frac{1}{z + i \lvl \supop{Q}} \op{\rho}(0) - i \frac{z}{z + i \lvl \supop{Q}} \lvl \supop{P} \op{\widetilde{\rho}}(z) \\
                                            &= - i \lvl \supop{Q} \frac{1}{z + i \lvl \supop{Q}} \op{\rho}(0) - i \left( \supopI - i \lvl \supop{Q} \frac{1}{z + i \lvl \supop{Q}} \right) \lvl \supop{P} \op{\widetilde{\rho}}(z) \\
                                            &= -i \lvl \supop{P} \op{\widetilde{\rho}}(z) - i \lvl \supop{Q} \frac{1}{z + i \lvl \supop{Q}} \op{\rho}(0) - \lvl \supop{Q} \frac{1}{z + i \lvl \supop{Q}} \lvl \supop{P} \op{\widetilde{\rho}}(z).
\end{align*}
Finally, this equation can be partially closed by applying $\supop{P}$ on both sides,
\begin{eqnarray}
\label{eq:SM-nakajima-zwanzig-laplace}
  \supop{P} \left( z \op{\widetilde{\rho}}(z) - \op{\rho}(0) \right) = -i \left( \supop{P} \lvl \supop{P} \right) \supop{P} \op{\widetilde{\rho}}(z) - i \supop{P} \lvl \supop{Q} \frac{1}{z + i \lvl \supop{Q}} \op{\rho}(0) - \supop{P} \lvl \supop{Q} \frac{1}{z + i \lvl \supop{Q}} \lvl \supop{P} \op{\widetilde{\rho}}(z).
\end{eqnarray}
Transforming back into the time domain, and making judicious use of the identities $\supop{Q}f(\lvl \supop{Q})\supop{Q} = \supop{Q}f(\supop{Q} \lvl \supop{Q})\supop{Q} = f(\supop{Q}\lvl )\supop{Q}$ for some function $f$ defined in terms of a power series, we arrive at
\begin{align}
  \label{eq:SM-nakajima-zwanzig}
  \frac{d}{dt} \supop{P} \op{\rho}(t) &= -i \left( \supop{P} \lvl \supop{P} \right) \supop{P} \op{\rho}(t) - i \supop{P} \lvl e^{-i t \supop{Q} \lvl} \supop{Q} \op{\rho}(0) - \int_0^t d\tau \, \left( \supop{P} \lvl e^{-i \tau \supop{Q} \lvl} \supop{Q} \lvl \supop{P} \right) \supop{P} \op{\rho}(t - \tau).
\end{align}
This is the Nakajima-Zwanzig equation governing the dynamics of a subset $\supop{P}\op{\rho}$ of all the degrees of freedom.

Note that because this derivation works in Laplace space, the same steps will hold for a general dynamics described by
\begin{align}
\begin{split}
    z \op{\widetilde{\rho}}(z) - \op{\rho}(0) &= -i \widetilde{\lvl}_{\text{eff}}(z) \op{\widetilde{\rho}}(z) + \op{\widetilde{\theta}}(z) \\
    \widetilde{\lvl}_{\text{eff}}(z) &= \lvl + \widetilde{\supop{K}}(z),
\end{split}
\end{align}
which extends the Nakajima-Zwanzig approach to open quantum systems as well.
Interestingly, one is therefore able to recursively iterate Nakajima-Zwanzig reduced dynamics, in a similar vein to how one might iterate decimation procedures in the renormalization group.

\section{\label{sec:SM-constraints}Structurally distinct generalized quantum master equations due to dynamical constraints}

Take a finite dimensional Hilbert space of $D$ number of states, which includes bath degrees of freedom.
Suppose there are $N$ explicitly identifiable conserved quantities in the dynamics, i.e.\ there are $N-1$ linearly-independent traceless Hermitian operators $\op{C}^{(k)}$ such that for some initial condition $\op{\rho}(0)$, the following holds at all times:
\begin{align}
    \Tr \big\{ \op{C}^{(k)} \op{\rho}(t) \big\} = q_k, \qquad \text{for $k = 2, \ldots, N.$}
\end{align}
In addition, let 
\begin{align}
    \op{C}^{(1)} = \frac{1}{\sqrt{D}} \op{I}.
\end{align}
Without loss of generality, let us assume that these operators form an orthonormal set with respect to the Hilbert-Schmidt norm, 
\begin{align}
    \Tr \bigg\{ \left( \op{C}^{(k)} \right)^\dagger \op{C}^{(k')} \bigg\} = \delta_{k,k'} \qquad \text{for all $k, k' = 1,\ldots, N$}.
\end{align}
This translates into $N$ linear constraints for the matrix elements of $\op{\rho}(t)$:
\begin{align}
    q_k &= \sum_{m=1}^D \sum_{n=1}^D C^{(k)}_{nm} \,\, \rho_{mn}(t), \qquad C^{(k)}_{nm} \equiv \langle n | \op{C}^{(k)} |m \rangle.
\end{align}
%There is no restriction on whether this space includes ``bath'' degrees of freedom, only that the full density matrix evolves according to a Liouville equation.
For simplicity, let us again consider a single distinguished observable, 
\begin{align}
    \sigma_{0}(t) &\equiv \Tr \big\{ (|0\rangle\langle 0| \otimes \op{I}_B) \op{\rho}(t) \big\}
\end{align}
for which we will attempt to derive generalized master equations.
We will assume that the operator $|0\rangle\langle 0| \otimes \op{I}_B$ is linearly independent from the set of $N$ conserved operators $\op{C}^{k}$.
One step of the Gram-Schmidt procedure allows us to orthogonalize the distinguished operator $|0\rangle\langle 0| \otimes \op{I}_B$ with respect to $\{\op{C}^{(k)}\}$ to give
\begin{align}
    \op{O} \equiv \frac{\left(|0\rangle\langle 0| \otimes \op{I}_B\right) - \sum_{k=1}^N \op{C}^{(k)} \Tr\bigg\{ (|0\rangle\langle 0| \otimes \op{I}_B) \op{C}^{(k)} \bigg\}}{\mathcal{N}},
\end{align}
where $\mathcal{N}$ is a normalization factor.
The remaining $D^2 - (N+1)$ components of the full density matrix $\op{\rho}$ can be orthonormalized in the same way, which we will call $r_1, \ldots, r_{D^2 - N - 1}$. 
If we vectorize the full density matrix and turn superoperators into matrices,
\begin{align}
    \op{\rho}(t) \Longrightarrow \vect{\rho}(t) = 
    \begin{pmatrix}
      \rho_{11}(t) \\ \vdots \\ \rho_{DD}(t) \\ \rho_{12}(t) \\ \vdots \\ \rho_{D,D-1}(t) 
    \end{pmatrix}, \qquad \lvl \Longrightarrow \mat{L}
\end{align}
then the orthonormal set of operators constructed above defines a \emph{time-independent} rotation matrix $\mat{R}$ such that
\begin{align}
    \label{eq:SM-rotated-rho-vector}
    \mat{R} \vect{\rho}(t) &= \begin{pmatrix}
      \frac{1}{\mathcal{N}} \left[ \sigma_{0}(t) - \sum_{k=1}^N q_k \Tr\bigg\{ (|0\rangle\langle 0| \otimes \op{I}_B) \op{C}^{(k)} \bigg\} \right] \\ q_1 \\ \vdots \\ q_N \\ r_1(t) \\ \vdots \\ r_{D^2 - N - 1}(t)
    \end{pmatrix}.
\end{align}
Because the basis-change is time-independent, the Liouville equation (in matrix form) is transformed as
\begin{align}
    \frac{d}{dt} \vect{\rho}(t) = -i \mat{L} \vect{\rho}(t) \qquad \Longrightarrow \qquad \frac{d}{dt} \mat{R} \vect{\rho}(t) &= -i \mat{R} \mat{L} \mat{R}^\dagger \,\, \mat{R} \vect{\rho}(t) \\
    &\equiv -i \mat{L}' \,\, \mat{R} \vect{\rho}(t). \nonumber
\end{align}
At this point we can project on to the distinguished variable using the projectors
\begin{align}
    \mat{P}^0 = 
    \begin{pmatrix}
      1 &   &        & \\
        & 0 &        & \\
        &   & \ddots & \\
        &   &        & 0
    \end{pmatrix} \qquad \text{and} \qquad 
    \mat{Q}^0 = 
    \begin{pmatrix}
      0 &   &        & \\
        & 1 &        & \\
        &   & \ddots & \\
        &   &        & 1
    \end{pmatrix},
\end{align}
which results in the Nakajima-Zwanzig equation
\begin{align}
  \frac{d}{dt} \mat{P}^0 \mat{R} \vect{\rho}(t) &= -i \left( \mat{P}^0 \mat{L}' \mat{P}^0 \right) \mat{P}^0 \mat{R} \vect{\rho}(t) \nonumber \\
  &\quad - i \mat{P}^0 \mat{L}' \,\, \exp(-i t \mat{Q}^0 \mat{L}') \,\, \mat{Q}^0 \mat{R} \vect{\rho}(0) \\
  &\quad - \int_0^t d\tau \, \bigg\{ \mat{P}^0 \mat{L}' \,\, \exp(-i \tau \mat{Q}^0 \mat{L}') \,\, \mat{Q}^0 \mat{L}' \mat{P}^0 \bigg\} \mat{P}^0 \mat{R} \vect{\rho}(t - \tau). \nonumber
\end{align}
This turns into a scalar equation after one projects on to the distinguished component on both sides. 
Because the value of the distinguished observable is real, the first term on the right hand side must be zero and one is generally left with
\begin{align}
    \frac{d}{dt} \sigma_0(t) &= \mathcal{N} \theta(t) - \int_0^t d\tau \,\, K(\tau) \left[ \sigma_0(t-\tau) - \sum_{k=1}^N q_k \Tr\bigg\{ (|0\rangle\langle 0| \otimes \op{I}_B) \op{C}^{(k)} \bigg\} \right], \\
    \intertext{where}
    \theta(t) &= \begin{pmatrix} 1 & 0 & \cdots & 0 \end{pmatrix} \bigg( - i \mat{P}^0 \mat{L}' \,\, \exp(-i t \mat{Q}^0 \mat{L}') \,\, \mat{Q}^0 \mat{R} \vect{\rho}(0) \bigg) \\
    K(\tau) &= \begin{pmatrix} 1 & 0 & \cdots & 0 \end{pmatrix} \bigg( \mat{P}^0 \mat{L}' \,\, \exp(-i \tau \mat{Q}^0 \mat{L}') \,\, \mat{Q}^0 \mat{L}' \mat{P}^0 \bigg) \begin{pmatrix} 1 \\ 0 \\ \vdots \\ 0 \end{pmatrix}.
\end{align}
Observe that the inhomogeneity $\theta(t)$ in general \emph{cannot} be set to zero even if we assume that the full density matrix is initially factorized, $\op{\rho}(0) = \op{\rho}_S \otimes \op{\rho}_B$.
This is because $\mat{Q}^0$ will pick out the conserved quantities $q_k$ in \nnref{eq:SM-rotated-rho-vector}{Eq.~(}{)}, which are generally nonzero.
Contrast this with the Nakajima-Zwanzig equation projected on to the $|0\rangle\langle 0|$ component of the system's reduced density matrix, where the inhomogeneity can be set to zero under factorized initial conditions.
Thus the presence of inhomogeneous terms in some forms of the generalized master equation follows from a  requirement to explicitly satisfy dynamical constraints. 
This formalizes the intuition we sketched out in the main text.

\section{\label{sec:SM-dynamics-equivalence}Time-nonlocal relations between memory kernels of different projectors and equivalence of generated dynamics}

\subsubsection{Warm up: Two-level system}
We will follow the convention of the main text, defining for $n=0,1$ the projectors
\begin{align}
\supop{P}^n \cdots = (|n\rangle\langle n|) \Tr \bigg\{ (|n\rangle\langle n|) \cdots \bigg\},
\end{align}
and memory kernels
\begin{align}
    K_{m,n}(\tau; \supop{P}) &= \Tr \bigg\{ (|m\rangle\langle m|) \lvl e^{-i \tau \supop{Q} \lvl} \supop{Q} \lvl (|n\rangle\langle n|)  \bigg\}.
\end{align}
Taking an initial state $\op{\sigma}(0) = |0\rangle\langle 0|$, we have the following master equations for the two projection schemes,
\begin{align}
    \label{eq:SM-tls-two-projected-master-eqs}
    \begin{cases}
    \displaystyle
    \frac{d \sigma_0(t)}{d t} = - \int_0^t d\tau \,\, K_{0,0}(\tau; \supop{P}^0) \sigma_0(t - \tau), & \supop{P} = \supop{P}^0 \\
    \displaystyle
    \frac{d \sigma_0(t)}{d t} = - \int_0^t d\tau \,\, \bigg(K_{0,0}(\tau; \supop{P}^0 + \supop{P}^1) + K_{1,1}(\tau; \supop{P}^0 + \supop{P}^1)\bigg) \sigma_0(t - \tau) + \int_0^t d\tau \,\, K_{1,1}(\tau; \supop{P}^0 + \supop{P}^1), & \supop{P} = \supop{P}^0 + \supop{P}^1
    \end{cases}
\end{align}
The memory kernels can be written as a derivative of functions $F(\tau)$, where
\begin{align}
    F_{m,n}(\tau; \supop{P}) = i\Tr \bigg\{ (|m\rangle\langle m|) \lvl e^{-i \tau \supop{Q} \lvl} (|n\rangle\langle n|) \bigg\} \quad \Longrightarrow \quad K_{m,n}(\tau; \supop{P}) = \frac{d}{d\tau} F_{m,n}(\tau; \supop{P}).
\end{align}
Note that $F_{m,n}(0; \supop{P}^0) = F_{m,n}(0; \supop{P}^0 + \supop{P}^1) = 0$.

We now work in Laplace space for $F$ using the larger projector $\supop{P}^0 + \supop{P}^1$, which has a complementary projector $\supopI - \supop{P}^0 - \supop{P}^1 = \supop{Q}^0 - \supop{P}^1$:
\begin{align}
    \widetilde{F}_{n,n}(z; \supop{P}^0 + \supop{P}^1) &= i\Tr \bigg\{ (|n\rangle\langle n|) \lvl \frac{1}{z +i (\supop{Q}^0 - \supop{P}^1) \lvl} (|n\rangle\langle n|) \bigg\} \\
    &= i\Tr \bigg\{ (|n\rangle\langle n|) \lvl \frac{1}{z + i \supop{Q}^0 \lvl} (|n\rangle\langle n|) \bigg\} \nonumber \\ 
    &\quad + i^2 \Tr \bigg\{ (|n\rangle\langle n|) \lvl \frac{1}{z + i \supop{Q}^0 \lvl} (|1\rangle\langle 1|) \bigg\} \Tr \bigg\{ (|1\rangle\langle 1|) \lvl \frac{1}{z +i (\supop{Q}^0 - \supop{P}^1) \lvl} (|n\rangle\langle n|) \bigg\} \\
    \widetilde{F}_{n,n}(z; \supop{P}^0 + \supop{P}^1) &= \widetilde{F}_{n,n}(z; \supop{P}^0) + \widetilde{F}_{n,1}(z; \supop{P}^0) \widetilde{F}_{1,n}(z; \supop{P}^0 + \supop{P}^1),
\end{align}
where we have used Dyson's identity in obtaining the second equality.
By using the identity $F_{0,n} + F_{1,n} = 0$, we therefore have
\begin{align}
    \label{eq:SM-tls-f-f0-f1-laplace}
    \widetilde{F}_{n,n}(z; \supop{P}^0 + \supop{P}^1) &= \widetilde{F}_{n,n}(z; \supop{P}^0) + \widetilde{F}_{1,1}(z; \supop{P}^0) \widetilde{F}_{n,n}(z; \supop{P}^0 + \supop{P}^1). 
\end{align}
We arrive at \nnref{eq:f0-f-identity}{Eq.~(}{)} in the main text by transforming back to the time domain.

We can cast the right hand side of \nnref{eq:SM-tls-f-f0-f1-laplace}{Eq.~(}{)} solely in terms of $\widetilde{F}$'s for the projector $\supop{P} = \supop{P}^0 + \supop{P}^1$.
Taking $n=1$, we find
\begin{align}
    \widetilde{F}_{1,1}(z; \supop{P}^0) &= \frac{\widetilde{F}_{1,1}(z; \supop{P}^0 + \supop{P}^1)}{1 + \widetilde{F}_{1,1}(z; \supop{P}^0 + \supop{P}^1)}.
\end{align}
Inserting this into the above identity for $n=0$ yields
\begin{align}
    \widetilde{F}_{0,0}(z; \supop{P}^0) &=  \frac{1}{1 + \widetilde{F}_{1,1}(z; \supop{P}^0 + \supop{P}^1)} \widetilde{F}_{0,0}(z; \supop{P}^0 + \supop{P}^1). \label{eq:SM-tls-f-laplace-decomposition}
\end{align}
This relation allows us to directly prove equivalence of the dynamics from both forms of master equations \nnref{eq:SM-tls-two-projected-master-eqs}{Eq.~(}{)}.
Note that, in terms of the $F$ functions, the solutions to these equations for the initial condition $\sigma_0(0) = 1$ is given in the Laplace domain as
\begin{align}
    \widetilde{\sigma}_0(z; \supop{P}^0) &= \frac{1}{z} \left[ \frac{1}{1 + \widetilde{F}_{0,0}(z; \supop{P}^0)} \right] \label{eq:SM-tls-pop-p-laplace} \\
    \widetilde{\sigma}_0(z; \supop{P}^0 + \supop{P}^1) &= \frac{1}{z} \left[ \frac{1 + \widetilde{F}_{1,1}(z; \supop{P}^0 + \supop{P}^1)}{1 + \widetilde{F}_{0,0}(z; \supop{P}^0 + \supop{P}^1) + \widetilde{F}_{1,1}(z; \supop{P}^0 + \supop{P}^1)} \right]. \label{eq:SM-tls-pop-p0-p1-laplace}
\end{align}
Inserting \nnref{eq:SM-tls-f-laplace-decomposition}{Eq.~(}{)} into \nnref{eq:SM-tls-pop-p-laplace}{Eq.~(}{)} immediately yields \nnref{eq:SM-tls-pop-p0-p1-laplace}{Eq.~(}{)}, which implies that the dynamics generated by the two approaches are identical.
Note that this follows solely from the assumption that the projectors for the memory kernels are $\supop{P}^0$ and $\supop{P}^0 + \supop{P}^1$.

We observe that one can derive a direct relation between the memory kernels of the two projection schemes by using the derivative relation $K(\tau) = \partial_\tau F(\tau) \Longrightarrow z \widetilde{F}(z) = \widetilde{K}(z)$.
Doing so in \nnref{eq:SM-tls-f-laplace-decomposition}{Eq.~(}{)} gives
\begin{align}
    \widetilde{K}_{0,0}(z; \supop{P}^0) &=  z\frac{1}{z + \widetilde{K}_{1,1}(z; \supop{P}^0 + \supop{P}^1)} \widetilde{K}_{0,0}(z; \supop{P}^0 + \supop{P}^1),
\end{align}
which in the time domain can be written with the help of an auxiliary function $G(t)$ such that
\begin{align}
    \label{eq:SM-tls-k-k0-k1-relation}
    \begin{split}
        K_{0,0}(t; \supop{P}^0) = K_{0,0}(t&; \supop{P}^0 + \supop{P}^1) +\!\! \int_0^t \!\!\!d\tau \,\, K_{0,0}(t - \tau; \supop{P}^0 + \supop{P}^1) \frac{d G(\tau)}{d\tau} \\
        \frac{d G(t)}{d t} = - \int_0^t \!\!\!d\tau \,\, &K_{1,1}(t - \tau; \supop{P}^0 + \supop{P}^1) G(\tau), \qquad G(0) = 1.
    \end{split}
\end{align}
%which is exactly the relation \nnref{eq:SM-tls-k-k0-k1-laplace}{Eq.~(}{)}.

\subsubsection{General case}
The case where the ``system'' dof has $D$ number of states and is coupled to a ``bath'' is more involved, but the ideas are the same as those for the two-level system.
Therefore, we will only define the necessary notation here, and then give the final result analogous to \nnref{eq:SM-tls-f-laplace-decomposition}{Eq.~(}{)}.
We will name the distinguished component of the system's reduced density matrix to be the population of the $|0\rangle$ state. %, denoted by
% \begin{align}
%     \sigma_{00}(t) \equiv \Tr \left\{ (|0\rangle\langle 0| \otimes \op{I}_B) \op{\sigma}(t) \right\}.
% \end{align}
We will define the projectors on to the $|m\rangle\langle n|$ component of the system's reduced density matrix as
\begin{align}
    \supop{P}^{m n} \op{\sigma} &= \bigg( |m\rangle\langle n| \otimes \op{\rho}_B \bigg) \Tr \bigg\{ (|n\rangle\langle m| \otimes \op{I}_B) \op{\sigma} \bigg\}.
\end{align}
We will take a general projector on to $N \leq D^2$ number of matrix elements of the system's reduced density matrix,
\begin{align}
    \supop{P} &= \supop{P}^{00} + \sum_{(m n) \in S} \supop{P}^{m n} \\
    &\equiv \supop{P}^{00} + \supop{P}^{\{S\}} \\
    S &\in \bigg\{ (m n) \bigg| \, 0 \leq \ell, m \leq D-1 \bigg\}\bigg\backslash \bigg\{ (00) \bigg\}.
\end{align}
Thus we will project on to some subset of matrix elements of the system's reduced density matrix, taking care to include the population $|0\rangle\langle 0|$.

First, we shall rewrite the Nakajima-Zwanzig equation (\nnref{eq:SM-nakajima-zwanzig}{Eq.~(}{)}) using the generalized force superoperator
\begin{align}
    \supop{F}(t; \supop{P}) &= \supop{P} \lvl e^{-i t \supop{Q} \lvl} \supop{P} \quad \Longrightarrow \supop{K}(t; \supop{P}) = \frac{d}{dt} \supop{F}(t; \supop{P})
\end{align}
and integration by parts, assuming the initial condition is factorized such that $\supop{Q}\op{\rho}(0) = 0$:
\begin{align}
    \frac{d}{dt} \supop{P} \op{\rho}(t) &= -i \left( \supop{P} \lvl \supop{P} \right) \supop{P} \op{\rho}(t) - \int_0^t d\tau \, \supop{K}(\tau; \supop{P}) \supop{P}\op{\rho}(t - \tau) \nonumber \\
    &= -\cancel{i \left( \supop{P} \lvl \supop{P} \right) \supop{P} \op{\rho}(t)} - \big( \supop{F}(t; \supop{P}) \supop{P} \op{\rho}(0) - \cancel{\supop{F}(0; \supop{P}) \supop{P} \op{\rho}(t)} \big) - \int_0^t d\tau \, \supop{F}(\tau; \supop{P}) \supop{P} \,\, \Dot{\op{\rho}}(t - \tau) \nonumber \\
    &= - \supop{F}(t; \supop{P}) \,\, \supop{P} \op{\rho}(0) - \int_0^t d\tau \, \supop{F}(\tau; \supop{P}) \,\, \supop{P} \Dot{\op{\rho}}(t - \tau).
\end{align}
In Laplace space, we can formally write the solution,
\begin{align}
    \widetilde{\supop{P} \op{\rho}}(z) &= \frac{1}{z} \left( \supop{P}\op{\rho}(0) - \left( \supopI + \widetilde{\supop{F}}(z; \supop{P}) \right)^{-1} \widetilde{\supop{F}}(z; \supop{P}) \supop{P} \op{\rho}(0) 
    %- \left( \supopI + \widetilde{\supop{F}}(z) \right)^{-1} i \supop{P} \lvl \frac{1}{z + i \supop{Q} \lvl} \supop{Q} \op{\rho}(0) 
    \right) \\ 
    &= \frac{1}{z} \left( \supopI + \widetilde{\supop{F}}(z; \supop{P}) \right)^{-1} \supop{P} \op{\rho}(0). \label{eq:SM-laplace-reduced-dynamics}
\end{align}

% Without loss of generality, we first consider the projection solely on to the population of the $|0\rangle$ state of the system's reduced density matrix, i.e.\ $\supop{P} = \supop{P}^{00}$. 
% Using this form in \nnref{eq:SM-laplace-reduced-dynamics}{Eq.~(}{)} and looking only at the $|0\rangle\langle 0|$ component of both sides, the Laplace-transformed dynamics is
% \begin{align}
%     \widetilde{\sigma}_{00}(z) &= \frac{1}{z} \left( \frac{1}{1 + \widetilde{F}_{00,00}(z; \supop{P}^{00})} \sigma_{00}(0) 
%     %- i \frac{1}{1 + \widetilde{F}^{00}_{00,00}(z)} \Tr\Bigg\{ (|0\rangle\langle 0| \otimes \op{I}_B) \lvl \frac{1}{z + i \supop{Q}^{00} \lvl} \supop{Q}^{00} \op{\rho}(0) \Bigg\} 
%     \right) \nonumber \\
%     &= \frac{1}{z} \left( \frac{1}{1 + \widetilde{F}_{(00),(00)}(z; \supop{P}^{00})} \right) \label{eq:SM-laplace-p0-dynamics}\\
%     \intertext{And matrix elements of $\supop{F}$ are given by}
%     \widetilde{F}_{(mn),(m'n')}(z; \supop{P}) &= \Tr \Bigg\{ (|n\rangle\langle m| \otimes \op{I}_B) \lvl \frac{1}{z + i \supop{Q}\lvl} (|m'\rangle\langle n'| \otimes \op{\rho}_B) \Bigg\},
% \end{align}
% since the factorized initial condition $\supop{P}^{00}\op{\rho}(0) = \op{\rho}(0) \Longrightarrow \op{\rho}(0) = |0\rangle\langle 0| \otimes \op{\rho}_B$.

We can cast \nnref{eq:SM-laplace-reduced-dynamics}{Eq.~(}{)} in a more familiar form by vectorizing operators and turning superoperators into matrices.
For instance, 
\begin{align}
    \widetilde{\supop{F}}(z; \supop{P}) \,\, \supop{P}\op{\rho} &= \supop{P} \lvl \frac{1}{z + i \supop{Q} \lvl} \supop{P} \op{\rho} \\
    &\Downarrow \nonumber \\
    \Tr \Bigg\{ (|n\rangle\langle m| \otimes \op{I}_B) \widetilde{\supop{F}}(z; \supop{P}) \,\, \supop{P}\op{\rho} \Bigg\}  &\equiv \sum_{(m'n')\in S\cup\{(00)\}} \widetilde{F}_{(mn),(m'n')}(z; \supop{P}) \,\, \sigma_{(m'n')} \\
    &= \sum_{(m'n')\in S\cup\{(00)\}} \underbrace{i \Tr\Bigg\{ (|n\rangle\langle m| \otimes \op{I}_B) \lvl \frac{1}{z + i \supop{Q} \lvl} (|m'\rangle\langle n'| \otimes \op{\rho}_B) \Bigg\}}_{\equiv \widetilde{F}_{(mn),(m'n')}(z; \supop{P})} \underbrace{\Tr\Bigg\{ (|n'\rangle\langle m'| \otimes \op{I}_B) \op{\rho} \Bigg\}}_{\equiv \sigma_{(m'n')}} \\
%    &= \sum_{m'n'} \widetilde{F}_{(mn),(m'n')}(z) \sigma_{(m'n')} \\
    \widetilde{\supop{F}}(z; \supop{P}) \,\, \supop{P}\op{\rho} \quad &\Longleftrightarrow \quad \widetilde{\mat{F}}(z; \supop{P}) \vect{\sigma}.
\end{align}
%Note that $\widetilde{F}$ and $\widetilde{F}^{00}$ differ in that the former uses the (complement) projection superoperator $\supop{Q}$ while the latter uses $\supop{Q}^{00}$.
% By using Dyson's identity
% %and that $\supop{Q} = \supop{Q}^{00} - \sum_{(m n) \in S} \supop{P}^{m n}$
% , we find a relation between different $\widetilde{F}$'s:
% \begin{align}
%     \widetilde{F}_{(mn),(m'n')}(z; \supop{P}^{00} + \supop{P}^{\{S\}}) &= \widetilde{F}_{(mn),(m'n')}(z; \supop{P}^{00}) + \sum_{(\mu \nu)\in S} \widetilde{F}_{(mn),(\mu\nu)}(z; \supop{P}^{00}) \,\, \widetilde{F}_{(\mu\nu),(m'n')}(z; \supop{P}^{00} + \supop{P}^{\{S\}}) \\
%     \begin{split}
%     &= \widetilde{F}_{(mn),(m'n')}(z; \supop{P}^{00}) + \sum_{(\mu \nu)} \widetilde{F}_{(mn),(\mu\nu)}(z; \supop{P}^{00}) \widetilde{F}_{(\mu\nu),(m'n')}(z; \supop{P}^{00} + \supop{P}^{\{S\}}) \\
%     &\qquad\, - \widetilde{F}_{(mn),(\mu\nu)}(z; \supop{P}^{00}) \underbrace{\Bigg( \delta_{(\mu\nu),(00)} \delta_{(00),(\mu'\nu')} \Bigg)}_{\equiv M_{(\mu\nu),(\mu'\nu')}} \widetilde{F}_{(\mu'\nu'),(m'n')}(z; \supop{P}^{00} + \supop{P}^{\{S\}})
%     \end{split}
% \end{align}
% This expression can be compactly written in the matrix notation as
Following the same steps as in the previous section, the generalization of \nnref{eq:SM-tls-f-laplace-decomposition}{Eq.~(}{)} is
\begin{empheq}[box=\widefbox]{align}
\label{eq:SM-f0-f-matrix-identity}
    \widetilde{\mat{F}}(z; \supop{P}^{00}) &= \widetilde{\mat{F}}(z; \supop{P}^{00} \,\, + \,\, \supop{P}^{\{S\}}) \bigg[ 1 \,\, + \,\, \widetilde{\mat{F}}(z; \supop{P}^{00} + \supop{P}^{\{S\}}) \,\, - \,\, \mat{P}^{00} \, \widetilde{\mat{F}}(z; \supop{P}^{00} + \supop{P}^{\{S\}}) \bigg]^{-1},
\end{empheq}
with the matrix version of the projector $\supop{P}^{00}$ being,
\begin{align}
\bigg( \mat{P}^{00} \bigg)_{(mn),(m'n')} &= \delta_{(mn),(00)} \delta_{(00),(m'n')}.
\end{align}
Proving the equivalence of the generated dynamics for the $|0\rangle\langle 0|$ component of the system's reduced density matrix is easiest with the following form of the above identity:
\begin{align}
    \widetilde{\mat{F}}(z; \supop{P}^{00} + \supop{P}^{\{S\}}) &= \left[ 1 \,\,-\,\, \widetilde{\mat{F}}(z; \supop{P}^{00}) \,\, + \,\, \widetilde{\mat{F}}(z; \supop{P}^{00}) \, \mat{P}^{00}) \right]^{-1} \widetilde{\mat{F}}(z; \supop{P}^{00}).
\end{align}

\section{\label{sec:SM-tls}Details on two-level system example}

In the main text, we showed two methods of deriving master equstions and memory kernels. 
Here we shall fill in the details of their derivation, as well as to present another approach to derive two-term ``master equations'' with entirely different time-independent memory kernels.
At least for the almost pathologically simple example of the two-level system, this highlights the possibility of nonuniqueness in the memory kernels, even if the structures of the master equations are the same.

We start with the Hamiltonian,
\begin{align}
\op{H} &= \varepsilon \op{\tau}^z + \Delta \op{\tau}^x.
\end{align}
If we represent the density operator of the qubit as a matrix,
\begin{align}
\op{\sigma} &= \begin{pmatrix} \sigma_{11} & \sigma_{10} \\ \sigma_{01} & \sigma_{00} \end{pmatrix},
\end{align}
the equation of motion for the density matrix is
\begin{align}
  \frac{d}{dt} \op{\sigma} &= -i \comm{ \begin{pmatrix} \varepsilon & \Delta \\ \Delta & -\varepsilon \end{pmatrix}}{\begin{pmatrix} \sigma_{11} & \sigma_{10} \\ \sigma_{01} & \sigma_{00} \end{pmatrix}} \\
  &= \begin{pmatrix} i \Delta (\sigma_{10} - \sigma_{01})  & - i \Delta (\sigma_{00} - \sigma_{11}) - 2 i \varepsilon \sigma_{10} \\ i \Delta (\sigma_{00} - \sigma_{11}) + 2 i \varepsilon \sigma_{01} & -i \Delta (\sigma_{10} - \sigma_{01}) \end{pmatrix}. \label{eq:SM-tls-EOM}
\end{align}
Note that we will be working in more generality here than in the main text, which focused only on the diagonal components of the density matrix $\sigma_0(t) \equiv \sigma_{00}(t)$ and $\sigma_1(t) \equiv \sigma_{11}(t)$.

\subsection{No projection; reduction via two constraints}
Besides the unitarity constraint, $\sigma_{00}(t) + \sigma_{11}(t) = 1$, there are three independent dynamical quantities whose equations of motion are
\begin{align}
  \frac{d}{dt} (\sigma_{10} - \sigma_{01}) &= -2 i \Delta (\sigma_{00} - \sigma_{11}) - 2 i \varepsilon(\sigma_{10} + \sigma_{01}) \label{eq:SM-qubit-coherence-diff-eom} \\
  \frac{d}{dt} (\sigma_{10} + \sigma_{01}) &= -2 i \varepsilon (\sigma_{10} - \sigma_{01}) \\
  \frac{d}{dt} (\sigma_{00} - \sigma_{11}) &= -2 i \Delta (\sigma_{10} - \sigma_{01}) \label{eq:SM-qubit-pop-diff-eom}
\end{align}
where one immediately finds an additional conserved quantity corresponding to the energy $E = \Tr\op{H} \op{\sigma}(t)$,
\begin{align}
  E &= \varepsilon (\sigma_{11}(t) - \sigma_{00}(t)) + \Delta (\sigma_{10}(t) + \sigma_{01}(t)). \label{eq:SM-conserved-quantity}
\end{align}
From \nnref{eq:SM-conserved-quantity}{Eq.~(}{)} we must therefore have
\begin{align}
\sigma_{10}(t) + \sigma_{01}(t) &= -\frac{\varepsilon}{\Delta} (\sigma_{11}(t) - \sigma_{00}(t)) + \frac{E}{\Delta} \label{eq:SM-tls-energy-constraint} \\
E &= \varepsilon \bigg(\sigma_{11}(0) - \sigma_{00}(0)\bigg) + \Delta \bigg(\sigma_{10}(0) + \sigma_{01}(0)\bigg).
\end{align}

At this point there are two ways to reduce the dynamics further, each resulting in a different ``master equation.''
\subsubsection{Method 1}

We can remove the degree of freedom $\sigma_{10}(t) + \sigma_{01}(t)$ exactly in  \nnref{eq:SM-qubit-coherence-diff-eom}{Eq.~(}{)} using the dynamical constraint \nnref{eq:SM-tls-energy-constraint}{Eq.~(}{)} to give
\begin{align}
  \begin{split}
    \frac{d}{dt}(\sigma_{10}(t) - \sigma_{01}(t)) &= 2i \Delta (\sigma_{11}(t) - \sigma_{00}(t)) \\
    & \quad - 2i \varepsilon \left[ - \frac{\varepsilon}{\Delta} (\sigma_{11}(t) - \sigma_{00}(t)) + \frac{E}{\Delta} \right]
    \end{split} \\
\begin{split}
  &= i \frac{2 (\Delta^2 + \varepsilon^2)}{\Delta} (\sigma_{11}(t) - \sigma_{00}(t)) - 2i E \frac{\varepsilon}{\Delta}.
  \end{split}
\end{align}
This can be formally integrated as
\begin{align}
  \begin{split}
  \sigma_{10}(t) - \sigma_{01}(t) &= (\sigma_{10}(0) - \sigma_{01}(0)) + i \int\limits_0^t \frac{2 (\Delta^2 + \varepsilon^2)}{\Delta} (\sigma_{11}(\tau) - \sigma_{00}(\tau)) d\tau - i\int\limits_0^t 2 E \frac{\varepsilon}{\Delta} d\tau.
  \end{split}
\end{align}
This can be substituted into \nnref{eq:SM-qubit-pop-diff-eom}{Eq.~(}{)} to give
\begin{align}
  \begin{split}
    \frac{d}{dt} (\sigma_{00}(t) - \sigma_{11}(t)) &= - 2 i \Delta (\sigma_{10}(0) - \sigma_{01}(0)) \\
    &\quad + \int\limits_0^t 4 (\Delta^2 + \varepsilon^2) \, (\sigma_{11}(\tau) - \sigma_{00}(\tau)) d\tau - \int\limits_0^t 4 E \varepsilon d\tau.
    \end{split}
\end{align}

If we specialize to the case where $\sigma_{01}(0) = \sigma_{10}(0) = 0$ and use the unitarity condition $1 = \sigma_{00}(t) + \sigma_{11}(t)$, this can be rewritten as
\begin{align}
    \frac{d}{dt} (\sigma_{00}(t) - \sigma_{11}(t)) &= - \int\limits_0^t \bigg[ 4 (\Delta^2 + \varepsilon^2) + 4 E \varepsilon \bigg] \sigma_{00}(\tau) + \bigg[ - 4 (\Delta^2 + \varepsilon^2) + 4 E \varepsilon \bigg] \sigma_{11}(\tau) d\tau \\
    &\Downarrow \nonumber \\
    \frac{d}{dt} \sigma_{00}(t) &= - \int\limits_0^t \bigg[ 2(\Delta^2 + \varepsilon^2) + 2 E \varepsilon \bigg] \sigma_{00}(\tau) + \bigg[ - 2 (\Delta^2 + \varepsilon^2) + 2 E \varepsilon \bigg] \sigma_{11}(\tau) d\tau.
\end{align}
Thus using only dynamical constraints, we obtain memory kernels
\begin{empheq}[box=\widefbox]{align}
\label{eq:SM-tls-constant-memory}
\begin{split}
    K_{0,0}(\tau) &= \phantom{-}2(\Delta^2 + \varepsilon^2) + 2 E \frac{\varepsilon}{\Delta} \quad\, \Longrightarrow \quad K_{1,0}(\tau) = -2(\Delta^2 + \varepsilon^2) - 2 E \frac{\varepsilon}{\Delta} \\
    K_{0,1}(\tau) &= -2(\Delta^2 + \varepsilon^2) + 2 E \frac{\varepsilon}{\Delta} \quad\, \Longrightarrow \quad K_{1,1}(\tau) = \phantom{-}2(\Delta^2 + \varepsilon^2) - 2 E \frac{\varepsilon}{\Delta}.
\end{split}
\end{empheq}
We remark that even though these kernels generate the correct dynamics, they are somewhat pathological in that some terms in the master equation grow without bound over time.

For the initial condition $\op{\sigma}(0) = |0\rangle\langle 0| \Longrightarrow E = -\varepsilon$, we find that these memory kernels indeed satisfy \nnref{eq:SM-tls-k-k0-k1-relation}{Eq.~(}{)}, i.e.,%where $G(t) = 2 \Delta^2 \cos(2 \omega t)$ and $\omega = \sqrt{\varepsilon^2 + (1/2)\Delta^2}$.
\begin{align}
    \frac{d}{dt} G(t) &= -\int_0^t d\tau \left( 2 \Delta^2 + 4 \varepsilon^2 \right) G(\tau), \quad G(0) = 1 \\
    \Longrightarrow G(t) &= \cos(2 \omega t), \quad \omega = \sqrt{\varepsilon^2 + (1/2)\Delta^2} \\
    K_{0,0}(t; \supop{P}^0) &= \int_0^t d\tau \, \left( 2 \Delta^2 \right) \frac{d G(\tau)}{d\tau} \nonumber \\
    &= 2\Delta^2 \cos(2\omega t).
\end{align}

\subsubsection{Method 2}
Using the same dynamical equations, we can write the energy constraint as
\begin{align}
    \sigma_{10}(t) &= - \sigma_{01}(t) -\frac{\varepsilon}{\Delta} (\sigma_{11}(t) - \sigma_{00}(t)) + \frac{E}{\Delta}. %\\
    %\sigma_{11}(t) &= 1 - \sigma_{00}(t).
\end{align}
We shall isolate only the equations of motion for $\sigma_{00}(t)$ and $\sigma_{01}(t)$ in \nnref{eq:SM-tls-EOM}{Eq.~(}{)}.
Using the forms of the two constraints as written above, we have
\begin{align}
    \frac{d}{dt} \sigma_{00}(t) &= -i \Delta (\sigma_{10}(t) - \sigma_{01}(t)) \nonumber \\
                                &= i \Delta \left(2 \sigma_{01}(t) + \frac{\varepsilon}{\Delta} ( \sigma_{11}(t) - \sigma_{00}(t)) - \frac{E}{\Delta} \right) \\
    \frac{d}{dt} \sigma_{01}(t) &= 2 i \varepsilon \sigma_{01}(t) - i \Delta (\sigma_{11}(t) - \sigma_{00}(t)).
\end{align}
We can formally solve for $\sigma_{01}(t)$ in terms of $\sigma_{00}(t)$, which gives
\begin{align}
    \sigma_{01}(t) &= -i \int_0^t d\tau \,\, \Delta e^{2i \varepsilon (t - \tau)} \, \left(\sigma_{11}(\tau) - \sigma_{00}(\tau) \right).
\end{align}
Thus the equation for $\sigma_{00}(t)$ can be closed to give
\begin{align}
    \frac{d}{dt} \sigma_{00}(t) &= i\left(\varepsilon - E \right) + i \varepsilon \left(\sigma_{11}(\tau) - \sigma_{00}(\tau) \right) + \int_0^t d\tau \,\, 2 \Delta^2 e^{2i \varepsilon (t - \tau)} \left(\sigma_{11}(\tau) - \sigma_{00}(\tau) \right).
\end{align}
Reality of the population means that we can focus only on the real parts, which give the ``master equation,''
\begin{align}
    \frac{d}{dt} \sigma_{00}(t) &= - \int_0^t \kappa_{0,0}(\tau) \sigma_{00}(t - \tau) - \kappa_{0,1}(\tau) \sigma_{11}(t - \tau) \, d\tau,
\end{align}
which implies,
\begin{empheq}[box=\widefbox]{align}
    \begin{split}
    \kappa_{0,0}(\tau) &= \phantom{-}2 \Delta^2 \cos(2 \varepsilon \tau) \quad \Longrightarrow \quad \kappa_{1,0}(\tau) = -2 \Delta^2 \cos(2 \varepsilon \tau) \\
    \kappa_{0,1}(\tau) &= -2 \Delta^2 \cos(2 \varepsilon \tau) \quad \Longrightarrow \quad \kappa_{1,0}(\tau) = \phantom{-}2 \Delta^2 \cos(2 \varepsilon \tau).
    \end{split} \label{eq:SM-tls-oscillating-memory}
\end{empheq}

Once again, these memory kernels satisfy \nnref{eq:SM-tls-k-k0-k1-relation}{Eq.~(}{)} with 
\begin{align}
    G(t) &= \frac{2 \varepsilon^2 + \Delta^2 \cos(2 \omega t)}{2 \omega^2}.
\end{align}

% \label{eq:SM-qubit-p0-p1-master-eq}

\subsection{Nakajima-Zwanzig master equation}
\subsubsection{Projection on to $|0\rangle$ population}
We define projection superoperators ($n=0,1$) to have action
\begin{align}
  \supop{P}^n \op{\sigma} &= \bigg( |n\rangle\langle n| \bigg) \Tr \left\{ (|n\rangle\langle n|) \op{\sigma} \right\} \nonumber \\
  &= \lrangle{n | \op{\rho} | n} \bigg( |n\rangle\langle n| \bigg),
\end{align}
such that we isolate only the dynamics of $\sigma_{nn}(t)$.
This can be used in the Nakajima-Zwanzig equation,
\begin{align}
  \frac{d}{dt} \supop{P}^0 \op{\sigma}(t) &= -i \left( \supop{P}^0 \lvl \supop{P}^0 \right) \supop{P}^0 \op{\sigma}(t) - i \supop{P}^0 \lvl e^{-i t \supop{Q}^0 \lvl} \supop{Q}^0 \op{\sigma}(0) - \int_0^t d\tau \, \supop{K}(\tau; \supop{P}^0) \supop{P}^0 \op{\sigma}(t - \tau). \nonumber \tag*{(\ref*{eq:SM-nakajima-zwanzig})}
\end{align}
We see that the first term of \nnref{eq:SM-nakajima-zwanzig}{Eq.~(}{)} is zero:
\begin{align}
  \supop{P}^0 \lvl \supop{P}^0 \op{\sigma} &= \lrangle{0 | \op{\rho} | 0} \bigg( \supop{P}^0 \comm{\op{H}}{|0\rangle\langle 0|} \bigg) \\
                                     &= \lrangle{0 | \op{\rho} | 0} \left\langle 0 \middle| \comm{\op{H}}{|0\rangle\langle 0|} \middle| 0 \right\rangle \bigg( |0\rangle\langle 0| \bigg) \\
                                     &= 0,
\end{align}
where the last line follows from the vanishing of the expectation value of the commutator.
We choose the second term of \nnref{eq:SM-nakajima-zwanzig}{Eq.~(}{)} to be zero by setting $\supop{Q}^0 \op{\sigma}(0) = 0 \Longrightarrow \op{\sigma}(0) = |0\rangle\langle 0|$.
Thus with this projector, the Nakajima-Zwanzig equation \nnref{eq:SM-nakajima-zwanzig}{Eq.~(}{)} becomes
\begin{align}
  |0\rangle\langle 0| \left( \frac{d}{dt} \lrangle{0 | \op{\sigma}(t) | 0} \right) &= 
  %|0\rangle\langle 0| \left\langle 0 \middle| -i \lvl e^{-i t \supop{Q} \lvl} \supop{Q} \op{\rho}(0) \middle| 0 \right\rangle 
  - |0\rangle\langle 0| \int\limits_0^t d\tau \, \left\langle 0 \middle| \lvl e^{-i \tau \supop{Q} \lvl} \supop{Q} \lvl \supop{P} \op{\rho}(t - \tau)  \middle| 0 \right\rangle.
\end{align}
We consider only the $|0\rangle\langle 0|$ component in this equation (cf.\ \nnref{eq:p0-nzme}{Eq.~(}{)} in the main text):
\begin{align}
  \frac{d}{dt} \sigma_{00}(t) &= 
  %-i \left\langle 0 \middle| \comm{\op{H}}{e^{-i t \supop{Q} \lvl} \supop{Q} \op{\rho}(0)} \middle| 0 \right\rangle 
  - \int\limits_0^t d\tau \, \left\langle 0 \middle| \lvl e^{-i \tau \supop{Q} \lvl} \supop{Q} \lvl \supop{P} \op{\sigma}(t - \tau)  \middle| 0 \right\rangle \\
  &= 
  %\underbrace{-i \left\langle 0 \middle| \comm{\op{H}}{e^{-i t \supop{Q} \lvl} \supop{Q} \op{\rho}(0)} \middle| 0 \right\rangle}_{\equiv \theta(t)} 
  - \int\limits_0^t d\tau \, \underbrace{\bigg\langle 0 \bigg| \lvl e^{-i \tau \supop{Q} \lvl} \supop{Q} \lvl (|0\rangle\langle 0|) \bigg| 0 \bigg\rangle}_{\equiv K_{0,0}(\tau; \supop{P}^0)} \sigma_{00}(t - \tau). \label{eq:SM-tls-nzme}
\end{align}
In order to explicitly compute this memory kernel, we first note that it can be written as the derivative of a simpler dynamical quantity
\begin{align}
  F_{0,0}(t; \supop{P}^0) &= i \bigg\langle 0 \bigg| \lvl e^{-i t \supop{Q}^0 \lvl} (|0\rangle\langle 0|) \bigg| 0 \bigg\rangle \\
  K(t; \supop{P}^0) &= \frac{d}{dt} F(t; \supop{P}^0).
\end{align}
We proceed by vectorizing the density matrix
\begin{align}
\op{\sigma} = \begin{pmatrix} \sigma_{11} & \sigma_{10} \\ \sigma_{01} & \sigma_{00} \end{pmatrix} \Longrightarrow \begin{pmatrix} \sigma_{00} \\ \sigma_{01} \\ \sigma_{10} \\ \sigma_{11} \end{pmatrix}.
\end{align}
In this basis, the projection operator $\supop{Q}^0$ and the Liouvillian are respectively represented by
\begin{align}
  \supop{Q}^0 =
              \begin{pmatrix}
                0 &  &  &  \\
                 & 1 &  &  \\
                 &  & 1 &  \\
                 &  &  & 1 \\
              \end{pmatrix} 
              \qquad\qquad \text{and} \qquad\qquad
   -i \lvl =
              \begin{pmatrix}
                0 & i\Delta  & -i\Delta & 0 \\
                i\Delta & 2i\varepsilon & 0 & -i\Delta \\
                -i\Delta & 0  & -2i\varepsilon & i \Delta \\
                0 & -i\Delta & i\Delta & 0 \\
              \end{pmatrix}.
\end{align}
The projected evolution starting from $|0\rangle\langle 0|$ will give
\begin{align}
e^{-i t \supop{Q}^0 \lvl} \begin{pmatrix} 1 \\ 0 \\ 0 \\ 0 \end{pmatrix} &= \begin{pmatrix} 1 \\ - \sin(\omega t) \frac{\varepsilon \Delta \sin(\omega t) - i \Delta \omega \cos(\omega t)}{\omega^2} \\ - \sin(\omega t) \frac{\varepsilon \Delta \sin(\omega t) + i \Delta \omega \cos(\omega t)}{\omega^2} \\ \Delta^2 \sin^2(\omega t) / \omega^2 \end{pmatrix}, \quad \quad \text{ where }
\omega = \sqrt{\frac{\Delta^2}{2} + \varepsilon^2}.
\end{align}
We can now calculate explicitly $F_{0,0}(t; \supop{P}^0)$ and $K_{0,0}(t; \supop{P}^0)$ to be
\begin{empheq}[box=\widefbox]{align}
  F_{0,0}(t; \supop{P}^0) = \Delta^2 \frac{\sin(2 \omega t)}{\omega} \quad \Longrightarrow \quad K_{0,0}(t; \supop{P}^0) = 2 \Delta^2 \cos(2 \omega t).
\end{empheq}
Finally, we arrive at the Nakajima-Zwanzig equation governing only the dynamics of the $|0\rangle\langle 0|$ part of the density matrix, assuming that $\op{\sigma}(0) = |0\rangle\langle 0|$:
\begin{align}
\begin{split}
  \frac{d}{dt} \sigma_{00}(t) = - \int\limits_0^t 2 \Delta^2 \cos(2 \omega \tau) \, \sigma_{00}(t - \tau) \, d\tau.
\end{split} \label{eq:SM-initial-p0-polarized-nzme}
\end{align}

\subsubsection{Projection on to $|0\rangle$ and $|1\rangle$ populations and reduction by one constraint}
One can repeat the same procedures as in the previous section to calculate $K_{n,n}(t; \supop{P}^0 + \supop{P}^1)$ for $n=0,1$.
Doing so, one finds that
\begin{empheq}[box=\widefbox]{align}
    \begin{split}
        K_{0,0}(t; \supop{P}^0 + \supop{P}^1) &= \phantom{-}2 \Delta^2 \cos(2 \varepsilon t) \quad \Longrightarrow \quad K_{1,0}(t; \supop{P}^0 + \supop{P}^1) = -2 \Delta^2 \cos(2 \varepsilon t) \\
        K_{0,1}(t; \supop{P}^0 + \supop{P}^1) &= -2 \Delta^2 \cos(2 \varepsilon t) \quad \Longrightarrow \quad K_{1,1}(t; \supop{P}^0 + \supop{P}^1) = \phantom{-}2 \Delta^2 \cos(2 \varepsilon t).
    \end{split}
\end{empheq}
These coincide with the kernels derived from only imposing dynamical constraints, \nnref{eq:SM-tls-oscillating-memory}{Eq.~(}{)}, but differ from \nnref{eq:SM-tls-constant-memory}{Eq.~(}{)} even though they all generate the same dynamics.
We hypothesize that the drastically dissimilar memory kernels is a phenomenon specific only to this simple example, as there are almost as many dynamical constraints as there are degrees of freedom.
\end{document}